# Strain engineering in single-, bi- and tri-layer MoS$_2$, MoSe$_2$, WS$_2$ and WSe$_2$

Felix Carrascoso[1,†], Hao Li[1,†], Riccardo Frisenda[1] (*), Andres Castellanos-Gomez[1] (*)

1 Materials Science Factory, Instituto de Ciencia de Materiales de Madrid, Consejo Superior de Investigaciones Científicas, 28049 Madrid, Spain.

† These two authors have contributed equally to this work

Riccardo.frisenda@csic.es, Andres.castellanos@csic.es

**ABSTRACT**

Strain is a powerful tool to modify the optical properties of semiconducting transition metal dichalcogenides like MoS$_2$, MoSe$_2$, WS$_2$ and WSe$_2$. In this work we provide a thorough description of the technical details to perform uniaxial strain measurements on these two-dimensional semiconductors and we provide a straightforward calibration method to determine the amount of applied strain with high accuracy. We then employ reflectance spectroscopy to analyze the strain tunability of the electronic properties of single-, bi- and tri-layer MoS$_2$, MoSe$_2$, WS$_2$ and WSe$_2$. Finally, we quantify the flake-to-flake variability by analyzing 15 different single-layer MoS$_2$ flakes.

## Introduction

Strain engineering, the modification of the electric and optical of materials through deformations of their crystal lattice, has become a powerful route to tune the properties of two-dimensional (2D) materials[1, 2]. The lack of dangling bonds on the surface of 2D materials makes them very resilient to mechanical deformations, close to the ideal limit predicted by Griffith[3–6], making it possible to tune the properties of 2D materials in a broad range. Moreover, unlike in typical strain engineering experiments in three-dimensional materials (where the strain is induced by forcing the epitaxial growth of one material onto another one with a lattice mismatch) in 2D materials strain can be applied through several different methods and it can be varied spatially and in time. These facts make 2D materials particularly suitable for strain engineering experiments and have motivated several works in the last years[7–13].

However, despite of the interest on strain engineering in 2D materials, we have found a lack of important technical details (such as the description of the experimental setups needed to apply strain or the method used to calibrate the strain) in the literature. Moreover, we found that most works about strain engineering on semiconducting transition metal dichalcogenides deal with a single-material and (most of the time) on a single-thickness[7, 8]. A comprehensive experimental study measuring different TMDCs of different thicknesses with the same experimental technique, setup and conditions may help to better understand the role of the chemical composition and number of layers in the strain tunability of TMDCs.

This is the motivation of our work. We firstly provide all the technical details to build up a three-points bending experimental setup compatible with optical microscopes to perform optical spectroscopy measurements upon strain. We then introduce a method to directly calibrate the amount of strain applied during the bending and we compare the experimentally determined strain with that obtained with continuous-mechanics formulas for bending of beams (typically used to determine the strain in strain engineering experiments in the literature). Lastly, we use the three-points bending setup to perform micro-reflectance measurements in single-, bi-, and tri-layer (labeled as 1L, 2L and 3L hereafter) MoS$_2$, MoSe$_2$, WS$_2$ and WSe$_2$ with comparable experimental conditions.

# Experimental setup and calibration

## Three-points-bending setup

We have selected a three-points bending geometry because, unlike two-points-bending geometry, it allows to keep the position of the specimen under study almost unchanged during the strain loading/unloading cycles. Therefore, this geometry is particularly advantageous to combine strain engineering experiments with optical spectroscopy techniques as it avoids re-focusing and re-positioning during the measurements, which could introduce unwanted artefacts in the data.

Our design of a three-points bending straining setup is based on two identical Z manual micrometric lineal stages (MAZ-40-10, Optics Focus, see one in Figure 1a). These stages are supplemented with homebuilt parts to attach the cylinders (MS1R/M, Thorlabs) that will be used as the pivots of the three-points-bending setup (as shown in Figure 1b and 1c). Permanent magnets are glued on the base and the stages are then attached to a magnetic steel base (see the complete setup in Figure 1d). Figures 1e to 1h show the detailed blueprints to build up the homebuilt parts. The assembly of the whole system is very easy, and its final cost is below 200 €. Note that our design could be adapted to perform four-points-bending experiments as well as biaxial straining with slight modifications of the homebuilt parts.

## Calibration of the applied strain upon bending

In many strain engineering works the applied strain is not a directly measured quantity, but it is instead typically derived from continuum mechanics formulas that correlate the applied strain with experimental parameters like the displacement of the clamping/pivotal points (leading to a deflection of the beam-shape substrate). In the case of a three-points-bending system the relationship between the strain ($\epsilon$), the distance between the pivotal points ($L$), the defection of the substrate ($D$) and the thickness of the substrate ($t$) is given by (see Figure 2a):

$$\epsilon = \frac{6Dt}{L^2} \quad \quad \text{Equation (1)}$$

Interestingly, we found a scarce amount of works where this kind of expressions are experimentally validated. We have developed a method to experimentally determine the applied strain in our bending test apparatus to validate Eq. (1). We fabricate arrays of photoresist pillars on the surface of the flexible substrates later used for the strain experiments (see details in the Materials and methods section). Then the substrate is mounted on the bending system and placed under the objective of an optical microscope to inspect the sample. Optical images of the samples, acquired at different substrate deflection, allow for a direct determination of the applied strain.

Figure 2b shows two optical images of a polycarbonate (PC) sample with photoresist pillars, acquired at zero deflection and at 8.5 mm deflection (corresponding to 0% and 4% strain accordingly to Equation 1). The experimental strain value can be determined by measuring the distance between the pillars at zero deflection ($l_0$) and that at a given substrate deflection ($l$):

$$\varepsilon_{(D)} = \frac{l-l_0}{l_0} \quad \quad \text{Equation (2)}$$

According to Eq. (1) the applied strain depends, apart from the deflection of the substrate, on the thickness of the substrate and the distance between the pivotal points. In order to check the validity of Eq. (1) we have experimentally measured the strain for different substrates (polycarbonate PC, polypropylene PP and mylar) with different thicknesses (Figure 2c) and for different distances between the pivotal points (Figure 2d). The experimental results (data points) agrees very well with the strain calculated obtained with Equation (1) (lines). Figure 2e plots the measured strain *vs.* the strain calculated with Eq. (1) for datasets with different materials, thicknesses and pivotal points distance. The data closely follow a linear trend with slope = 1, thus validating Eq. (1) to determine the applied strain in our experimental setup.

Note that this calibration method provides an upper limit for the strain value. The actual strain on the 2D material, transferred on top of the flexible substrate, will depend on the efficiency of strain transfer. Strain transfer strongly depends on the Young's modulus of the flexible substrate, being higher for substrates with high Young's modulus.[14, 15] According to finite elements calculations substrates with a Young's modulus higher than 1 GPa should yield an almost perfect strain transfer.[15, 16] In this work we have tested different substrates, polypropylene (~1.5-2 GPa), polycarbonate (~2.5 GPa) and mylar (~4-5 GPa), finding similar results. (see Fig. S1, Fig. S2 and Fig. S3 in the Electronic Supplementary Material (ESM)).

## Materials and methods

### Isolation and identification of the 2D materials

$MoS_2$ bulk crystal was obtained from natural molybdenite mineral (Molly Hill mine, Quebec, Canada). $MoSe_2$, $WS_2$ and $WSe_2$ bulk crystals (synthetic) were purchased at HQ Graphene. The transition metal dichalcogenide bulk crystals were exfoliated with Nitto tape (Nitto SPV 224) and the resulting cleaved crystallites were transferred to the surface of a Gel-Film (Gel-Pak, WF ×4 6.0 mil) substrate by adhering the tape containing the crystallites and peeling-off slowly. Single-, bi- and tri-layer flakes are identified at first glance by inspecting the Gel-Film substrate with transmission mode optical microscopy (Motic BA 310 MET-T optical microscope). Quantitative analysis of the transmission mode optical images provides a first way to assess the number of layers [17, 18] that can be further double-checked by micro-reflectance spectroscopy.[19]

### Placement of 2D materials onto flexible substrates

Once a suitable flake is identified it is transferred into the center of a flexible substrate with an all-dry deterministic placement method.[20–22] In this work we have explored the use of polycarbonate (PC: thickness 250 μm and 750 μm), polypropylene (PP: thickness 185 μm) and mylar (thickness 100 μm) (see Fig. S1, Fig. S2 and Fig. S3 in the Electronic Supplementary Material (ESM)).

### Strain dependent differential reflectance spectroscopy

The flexible substrate with the flake to be studied is loaded into the three-points bending system and the whole system is mounted in an optical microscope system supplemented with a homebuilt micro-reflectance module based on a fiber-coupled CCD spectrometer (CCS200/M, Thorlabs) discussed in details in our previous work [23]. The flake under study is centered with respect to the central pivot by sliding the substrate while the sample is inspected under the microscope. Once the flake under study is aligned the strain engineering experiment can be carried out by acquiring differential reflectance spectra as a function of the applied strain. We decided to use differential reflectance instead of photoluminescence, commonly used in the literature to probe the strain effect on the optical properties of 2D semiconductors, as differential reflectance allows to easily resolve other excitonic features apart from the A exciton, even for multilayer flakes. See Fig. S4 in the Electronic Supplementary Material (ESM) for a comparison between photoluminescence and differential reflection spectra acquired on the same flake upon uniaxial strain. Moreover, it has been shown that differential reflectance provides similar information as compared with transmission or absorption spectroscopy.[19]

## Results

Figure 3a shows a summary of differential reflectance spectra, acquired at different strain levels for a single-layer $MoS_2$ flake (see inset in Figure 3b). The spectra show two prominent peak features at ~1.9 eV and ~2.0 eV that correspond to direct band gap transitions at the K point of the Brillouin zone that generates excitons labelled A and B in agreement with previous literature[19, 24, 25]. Both A and B excitons redshift upon increasing tensile strain. The measured spectra can be fitted to a sum of two Gaussian peaks with a smooth second order polynomial background to extract the position of the excitons, their full-width-at-half maximum (FWHM) and amplitudes. We have found that in some cases strains up to 2% (see Fig. S2 in the Electronic Supplementary Material (ESM)) can be applied before slippage but slippage typically occurs between 0.8% and 1.4% (see Fig. S3 in the Electronic Supplementary

Material (ESM) for a measurement of the strain cycling reproducibility). Although we do not have a complete microscopic understanding of the slippage process, we have qualitatively observed that large-area flakes with low density of defects or folds can sustain larger strain before slippage. Thus, we believe that the friction force, needed to displace a flake, is the main mechanism preventing its slippage and that when the applied stress overcomes the static friction force slippage occurs.

Figure 3b shows the A and B exciton energy values upon straining. They follow a linear trend whose slope determines the strain gauge factor, the amount of spectral shift per % of uniaxial strain, being -37 ± 1 meV/% and -34 ± 1 meV/% for the A and B excitons respectively. These values are in good agreement with those reported in the literature. For example, the A exciton gauge factor values for uniaxially strained 1L-$MoS_2$ in the literature span over a broad range between -3 meV/% and -125 meV/% with a median value of -37 meV/%. This large scattering in the reported gauge factor values could be caused by different sources: differences in the experimental conditions (i.e. different strain transfer between the different employed substrates), flake-to-flake variation and/or a wrong estimation of the strain transfer. In order to estimate the magnitude of the flake-to-flake variability in the span of reported gauge factors, we have studied 15 different 1L-$MoS_2$ flakes at different uniaxial strains up to 1.3% (see Fig. S5, Fig. S6 and Fig. S7 in the Electronic Supplementary Material (ESM)). Figure 4 shows a summary of the statistical information obtained after analyzing the 15 different datasets. Figure 4a shows a box plot with the statistical information of the A and B exciton energies at different strain levels. This representation shows the expected overall trend of redshifting of the A and B exciton energies upon straining with a relative low variability (50% of the datapoints show a scattering of <15 meV). Figure 4b and 4c represent the statistical variation of the amplitude and the FWHM of the A and B excitons respectively at different uniaxial strain values, showing no statistically relevant strain dependence. In order to determine the magnitude of flake-to-flake variation, and its contribution to explain the scattering in the gauge factor values reported in the literature, Figure 4d shows a box plot comparing the statistical information of the 42 1L-$MoS_2$ flakes reported in the literature (measured on different substrates, under different experimental conditions, etc.) with that of our 15 1L-$MoS_2$ (measured at identical experimental conditions). The direct comparison between the two boxes indicates that the flake-to-flake variation itself cannot explain the large scattering reported in the literature. We can thus conclude that the differences in the experimental conditions and/or strain calibration in the different reported works are the main sources of scattering between reported values. Therefore, one should be careful when comparing gauge factors from different experimental sources, trying to compare sources using as similar as possible experimental conditions.

We have further repeated the straining experiments shown in Figure 3 for 1L, 2L and 3L $MoS_2$, $MoSe_2$, $WS_2$ and $WSe_2$ flakes (see Fig. S8 to S18 in the Electronic Supplementary Material (ESM)). The corresponding datasets can be found in the Supporting Information. Table 1 summarizes the respective gauge factors for these datasets.

The obtained values for 1L and 2L flakes are compatible with those reported in the literature (summarized in Tables 2, 3, 4 and 5). Note that as far as we know, this is the first report for uniaxial strain tuning for 3L flakes of these transition metal dichalcogenides in the literature. Because of the flake-to-flake statistical variability inferred from the analysis of 15 1L-$MoS_2$ flakes we can only claim that the thickness or chemical composition dependence of the gauge factors seem to be smaller or at least comparable to the flake-to-flake variability.

## Conclusions

We reported a comprehensive study of the effect of uniaxial strain on the optical properties of several semiconducting transition metal dichalcogenides. We describe all the technical details to assemble a uniaxial strain tuning setup based on a three-points-bending system that is very suitable for its use in optical spectroscopy experiments. We also introduced a very straightforward method to calibrate the amount of strain induced during the bending cycles. We used our system to study the strain tunable excitonic features in 1L $MoS_2$ and we analyzed the flake-to-flake variability analyzing 15 different single-layer flakes. We further studied 1L, 2L and 3L $MoS_2$, $MoSe_2$, $WS_2$ and $WSe_2$, reporting (as far as we know) the first results on trilayer transition metal dichalcogenide flakes.

## Acknowledgements

This project has received funding from the European Research Council (ERC) under the European Union's Horizon 2020 research and innovation programme (grant agreement n° 755655, ERC-StG 2017 project 2D-TOPSENSE). R.F. acknowledges the support from the Spanish Ministry of Economy, Industry and Competitiveness through a Juan de la Cierva-formación fellowship 2017 FJCI-2017-32919. LH acknowledges the grant from China Scholarship Council (CSC) under No. 201907040070.

**Electronic Supplementary Material:** Datasets for $MoS_2$ acquired on different flexible substrates, reproducibility of the strain cycles in 3L-$MoS_2$, comparison between photoluminescence and differential reflectance upon strain, statistical analysis of the flake-to-flake variation in 1L $MoS_2$, datasets for 2L and 3L $MoS_2$, datasets for 1L, 2L and 3L $MoSe_2$, datasets for 1L, 2L and 3L $WS_2$ and datasets for 1L, 2L and 3L $WSe_2$ are available in the online version of this article at http://dx.doi.org/10.1007/s12274-***-****-* (automatically inserted by the publisher).

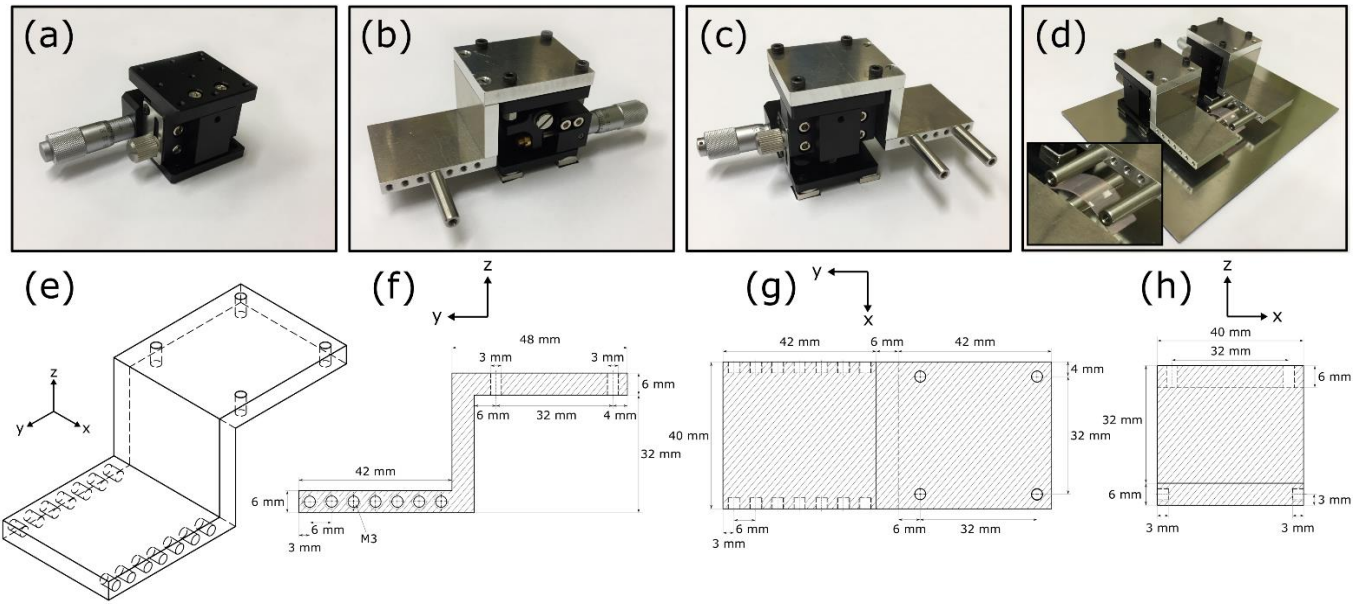

**Figure 1.** Homebuilt three-points-bending setup for uniaxial strain experiments on 2D materials. (a) Picture of one of the manual Z linear stages used for the assembly of the setup. (b) and (c) Pictures of the two manual Z stages after attaching the homemade parts. (d) Picture of the finished setup with a flexible substrate under a strain test (inset). (e) to (h) Blueprints of the homemade parts.

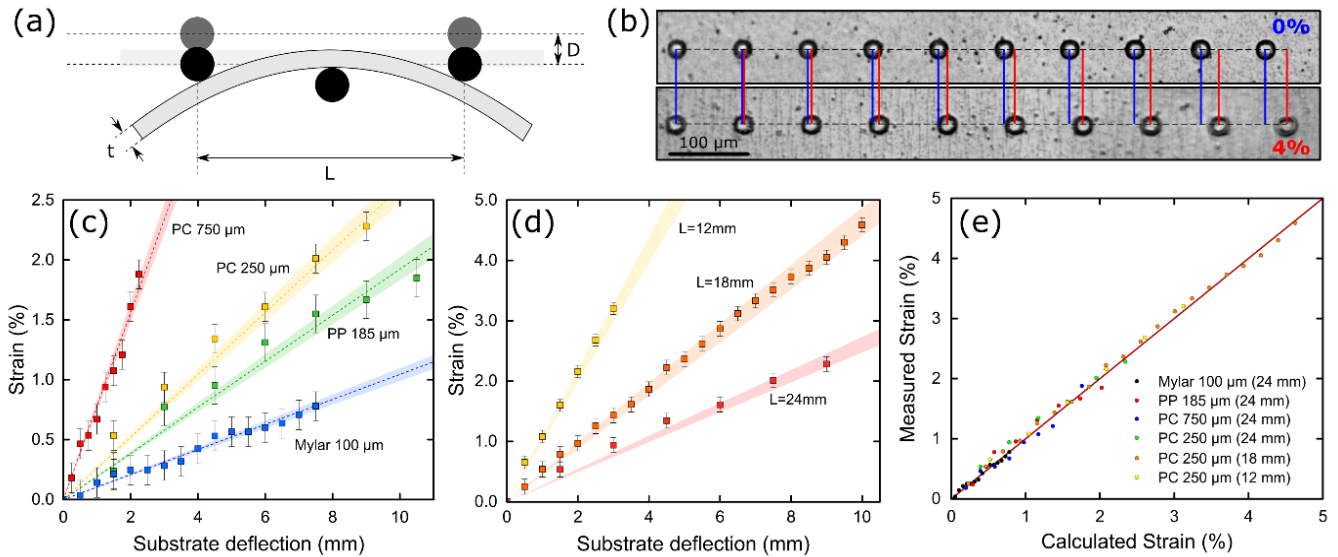

**Figure 2.** Calibration of the three-points-bending setup. (a) Scheme of the three-points-bending experiment indicating the relevant experimental magnitudes: distance between pivotal points ($L$), thickness of the substrate ($t$) and deflection of the substrate ($D$). (b) Optical microscopy images of photoresist micropillars patterned on the surface of a polycarbonate (PC) substrate 250 µm thick before and after bending the substrate. The displacement of the pillars provides a direct measurement of the strain. (c) Measured strain as a function of the substrate deflection (D) for substrates with different thickness (datapoints) and the calculated values using Eq. (1) (lines, the shadowed area around indicates the uncertainty of the calculated value). (d) Measured strain as a function of the substrate deflection (D) for the same substrate while varying the pivotal distance (datapoints) and the calculated values using Eq. (1) (lines, the shadowed area around indicates the uncertainty of the calculated value). (e) Relationship between the measured strain values and those calculated using Eq. (1) for different substrates with different pivotal distances. A line with slope = 1 is displayed to indicate the small deviations with respect to the perfect agreement.

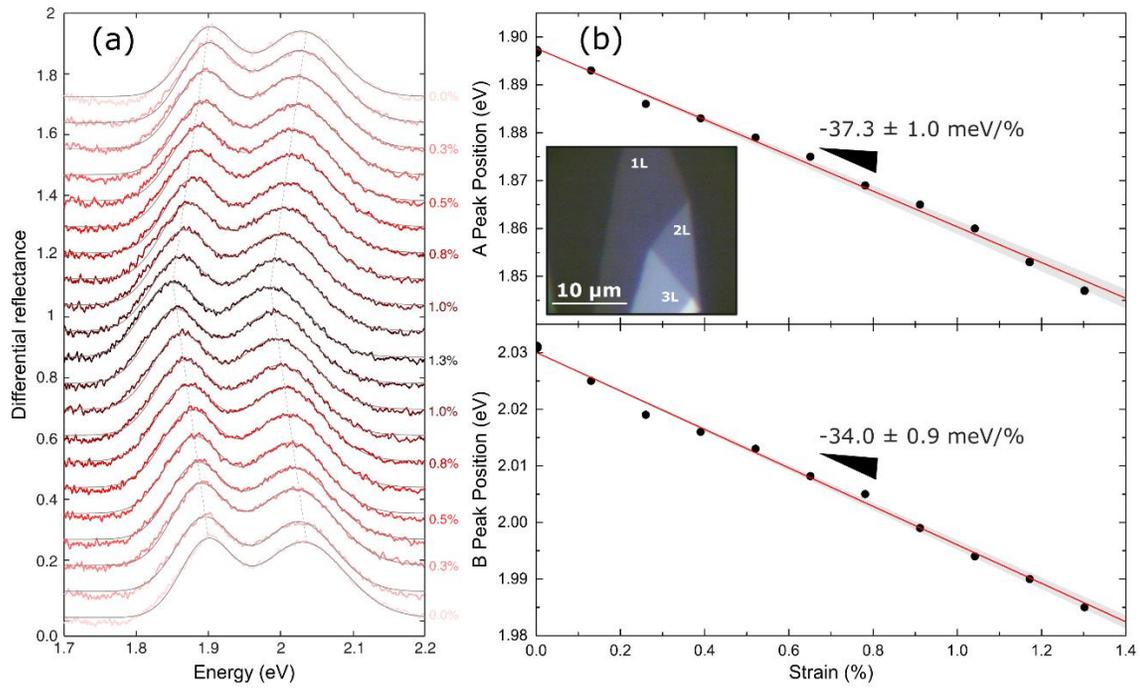

**Figure 3.** Strain tunable differential reflectance in 1L-MoS$_2$. (a) Differential reflectance spectra acquired at different uniaxial strain levels, up to 1.3%. (b) A and B exciton energy values as a function of the applied uniaxial strain. A linear fit is employed to extract the gauge factor, the spectral shift per % of uniaxial strain, which is indicated in each panel.

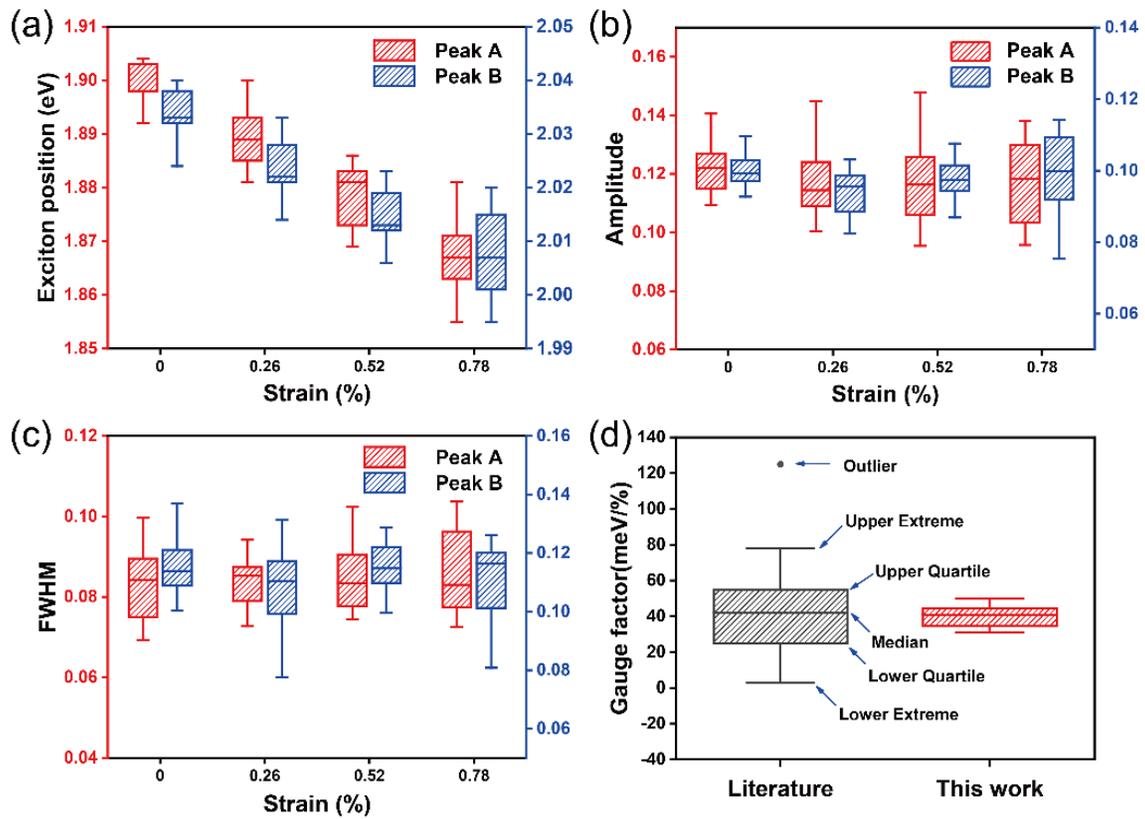

**Figure 4.** Statistical flake-to-flake variation in 15 different 1L-MoS$_2$. (a) Box plot representation of the A and B exciton energy values at different uniaxial strain values. (b) and (c) Box plot representation of the amplitude and FWHM of the A and B excitons at different strain values. (d) Comparison between the statistical variability reported in the literature for 42 1L-MoS$_2$ flakes from different studies and our 15 1L-MoS$_2$ flakes. Note: in panel (d) we include an explanation of the different parts displayed in the box plot representation.

**Table 1.** Summary of the experimental gauge factors for the different excitons measured in 1L, 2L and 3L MoS$_2$, MoSe$_2$, WS$_2$ and WSe$_2$. * Analysis of 15 different 1L MoS$_2$ flakes.

| Material | # layers | Gauge factor (meV/%) | |
|---|---|---|---|
| | | A | B |
| MoS$_2$ | 1L | -37 ± 1 | -34 ± 1 |
| | 1L* | -31.4 to -50.6 | -18.6 to -37.9 |
| | 2L | -38 ± 3 | (B) -42 ± 2 / (IL) -43 ± 1 |
| | 3L | -38 ± 1 | -36 ± 1 |
| MoSe$_2$ | 1L | -41 ± 3 | -22 ± 1 |
| | 2L | -28 ± 1 | -19 ± 7 |
| | 3L | -20 ± 2 | -25 ± 2 |
| WS$_2$ | 1L | -44 ± 2 | -- |
| | 2L | -31 ± 3 | -38 ± 7 |
| | 3L | -40 ± 4 | -46 ± 7 |
| WSe$_2$ | 1L | 57 ± 4 | -41 ± 4 |
| | 2L | -50.7 ± 6 | -61 ± 6 |
| | 3L | -28 ± 5 | -16 ± 2 |

**Table 2.** Summary of the reported experimental gauge factors in the literature for uniaxially strained MoS$_2$. If not indicated in the 'Material' column the material is mechanically exfoliated. * Statistical analysis over 28 different flakes. ** Data obtained in a polyvynilacetate (PVA) encapsulated sample. *** Data obtained in a sample fabricated onto highly compliant polydimethylsiloxane (PDMS).

| Work | Material | Experiment | Gauge factor (meV/%) |
|---|---|---|---|
| Conley *et al*. [7] | 1L MoS$_2$ | PL (A exciton) | -45 ± 7 |

|  | 1L MoS$_2$* | PL (A exciton) | -5 to -75 |
|---|---|---|---|
|  | 2L MoS$_2$ | PL (A exciton) | -53 ± 10 |
|  |  | PL (I exciton) | -129 ± 20 |
| He et al. [8] | 1L MoS$_2$ | PL & Abs (A exciton) | -64 ± 5 |
|  |  | PL & Abs (B exciton) | -68 ± 5 |
|  | 2L MoS$_2$ | PL (A exciton) | -48 ± 5 |
|  |  | Abs (A exciton) | -71 ± 5 |
|  |  | PL & Abs (B exciton) | -67 ± 5 |
|  |  | PL (I exciton) | -77 ± 5 |
| Zhu et al. [12] | 1L MoS$_2$ | PL (A exciton) | -48 |
|  | 2L MoS$_2$ | PL (A exciton) | -46 |
|  |  | PL (I exciton) | -86 |
| Niehues et al. [26] | 1L MoS$_2$ | PL (A exciton) | -28 ± 1 |
|  |  | Abs (A exciton) | -42 ± 2 |
| Niehues et al. [27] | 1L MoS$_2$ | Abs (A exciton) | -42 ± 2 |
|  |  | Abs (B exciton) | -38 ± 4 |
|  | 1L MoS$_2$ CVD | Abs (A exciton) | -42 ± 2 |
|  |  | Abs (B exciton) | -43 ± 4 |
| Niehues et al. [28] | 2L MoS$_2$ | Abs (A exciton) | -49 ± 1 |
|  |  | Abs (B exciton) | -49 ± 1 |
|  |  | Abs (IL exciton) | -47 ± 2 |
| Christopher et al. [29] | 1L MoS$_2$ CVD | PL (A exciton) | -38 ± 1 |
|  |  | PL (A$^-$ exciton) | -43 ± 1 |
|  |  | PL (B exciton) | -50 ± 2 |
| Zhiwei Li et al. [30] | 1L MoS$_2$** | PL (A exciton) | -125 |
|  | 1L MoS$_2$ | PL (A exciton) | -61 |
| Xin He [31] | 1L MoS$_2$ | PL (A exciton) | -56 |
| Zheng Liu [32] | 1L MoS$_2$*** CVD | PL (A exciton) | -3 |
| John et al. [33] | 1L MoS$_2$ | PL (A exciton) | -78 ± 4 |
|  | 2L MoS$_2$ | PL (A exciton) | -34 ± 3 |
|  |  | PL (I exciton) | -155 ± 11 |

**Table 3.** Summary of the reported experimental gauge factors in the literature for uniaxially strained MoSe$_2$.

| Work | Material | Experiment | Gauge factor (meV/%) |
|---|---|---|---|
| Island et al. [34] | 1L MoSe$_2$ | PL (A exciton) | -27 ± 2 |
| Niehues et al. [26] | 1L MoSe$_2$ | PL (A exciton) | -38 ± 2 |
|  |  | Aba (A exciton) | -34 ± 2 |
| Mennel et al. [35] | 1L MoSe$_2$ | PL (A exciton) | -55 ± 6 |

**Table 4.** Summary of the reported experimental gauge factors in the literature for uniaxially strained WS$_2$. If not indicated in the 'Material' column the material is mechanically exfoliated. * Data obtained in a polyvynilacetate (PVA) encapsulated sample. ** Data obtained in a sample fabricated onto highly compliant polydimethylsiloxane (PDMS).

| Work | Material | Experiment | Gauge factor (meV/%) |
|---|---|---|---|
| Wang et al. [13] | 1L WS$_2$ | PL (A exciton) | -11 |
| | | PL (A$^-$ exciton) | -11 |
| | | PL (I exciton) | -19 |
| Niehues et al. [26] | 1L WS$_2$ | PL (A exciton) | -50 ± 2 |
| | | Abs (A exciton) | -55 ± 2 |
| Zhiwei Li et al. [30] | 1L WS$_2$* CVD | PL (A exciton) | -43 |
| Qianhui Zhang [36] | 1L WS$_2$** CVD | PL (A exciton) | -1.3 |
| Xin He [31] | 1L WS$_2$ | PL (A exciton) | -43 |
| | | PL (A$^-$ exciton) | -46 |
| Mennel et al. [35] | 1L WS$_2$ | PL (A exciton) | -61.2 ± 3.8 |

**Table 5.** Summary of the reported experimental gauge factors in the literature for uniaxially strained WSe$_2$. If not indicated in the 'Material' column the material is mechanically exfoliated. * Data obtained in a polyvynilacetate (PVA) encapsulated sample.

| Work | Material | Experiment | Gauge factor (meV/%) |
|---|---|---|---|
| Schmidt et al. [37] | 1L WSe$_2$ | Abs (A exciton) | -54 ± 2 |
| | | Abs (B exciton) | -50 ± 3 |
| | | Abs (C exciton) | +17 ± 2 |
| | | Abs (D exciton) | -22 ± 2 |
| Niehues et al. [26] | 1L WSe$_2$ | PL & Abs (A exciton) | -49 ± 2 |
| Zhiwei Li et al. [30] | 1L WSe$_2$* | PL (A exciton) | -109 |
| | 1L WSe$_2$* CVD | PL (A exciton) | -53 |
| Desai [38] | 2L WSe$_2$ | PL (A exciton) | -45 |
| Mennel et al. [35] | 1L WSe$_2$ | PL (A exciton) | -53 ± 3 |
| Aslan et al. [39] | 1L WSe$_2$ | Abs (A exc.) | -48 |
| | | Abs (B exc.) | -40 |
| | | PL (A exc.) | -44 |
| Aslan et al. [40] | 2L WSe$_2$ | Reflectance (A exc.) | -52 |
| | | Reflectance (B exc.) | -48 |
| Tang et al. [41] | 2L WSe$_2$ | PL (A exciton) | -22.5 |
| | | PL (I exciton) | +20 |

# Electronic Supplementary Material

# Strain engineering in single-, bi- and tri-layer $MoS_2$, $MoSe_2$, $WS_2$ and $WSe_2$

Felix Carrascoso[1,†], Hao Li[1,†], Riccardo Frisenda[1] (*), Andres Castellanos-Gomez[1] (*)

1 Materials Science Factory, Instituto de Ciencia de Materiales de Madrid, Consejo Superior de Investigaciones Científicas, 28049 Madrid, Spain.

† These two authors have contributed equally to this work

**ELECTRONIC SUPPLEMENTARY MATERIAL CONTENT:**

**Datasets for $MoS_2$ acquired on different flexible substrates**

**Reproducibility of the strain cycles in 3L-$MoS_2$**

**Comparison between photoluminescence and differential reflectance in 1L $MoS_2$**

**Statistical analysis of the flake-to-flake variation in 1L $MoS_2$**

**Datasets for 2L and 3L $MoS_2$**

**Datasets for 1L, 2L and 3L $MoSe_2$**

**Datasets for 1L, 2L and 3L $WS_2$**

**Datasets for 1L, 2L and 3L $WSe_2$**

**DATASETS FOR MOS$_2$ ACQUIRED ON DIFFERENT FLEXIBLE SUBSTRATES:**

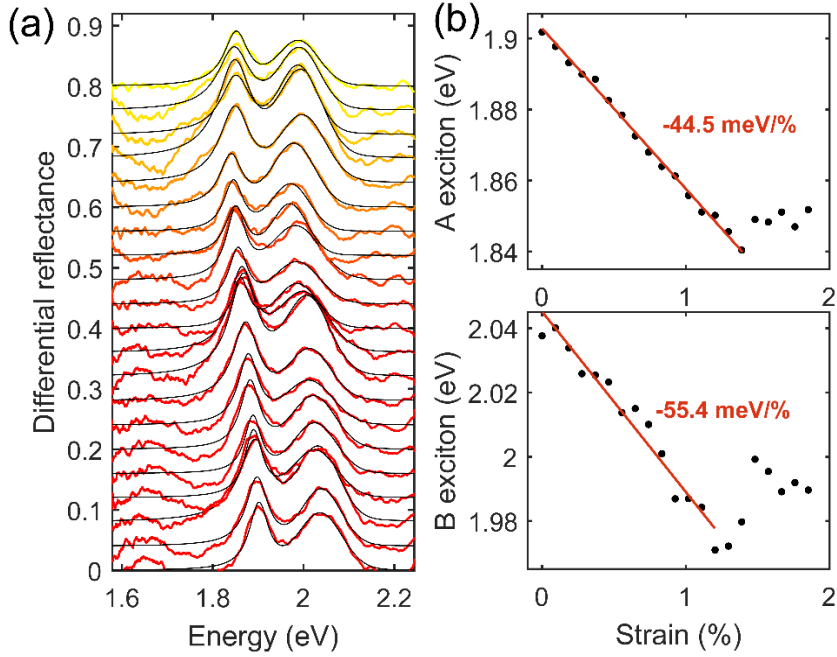

**Figure S1.** Strain tunable differential reflectance in 1L-MoS$_2$ on a mylar substrate (100 μm thick). (a) Differential reflectance spectra acquired at different uniaxial strain levels, slippage starts at ~1.3%. (b) A and B exciton energy values as a function of the applied uniaxial strain. A linear fit is employed to extract the gauge factor, the spectral shift per % of uniaxial strain, which is indicated in each panel.

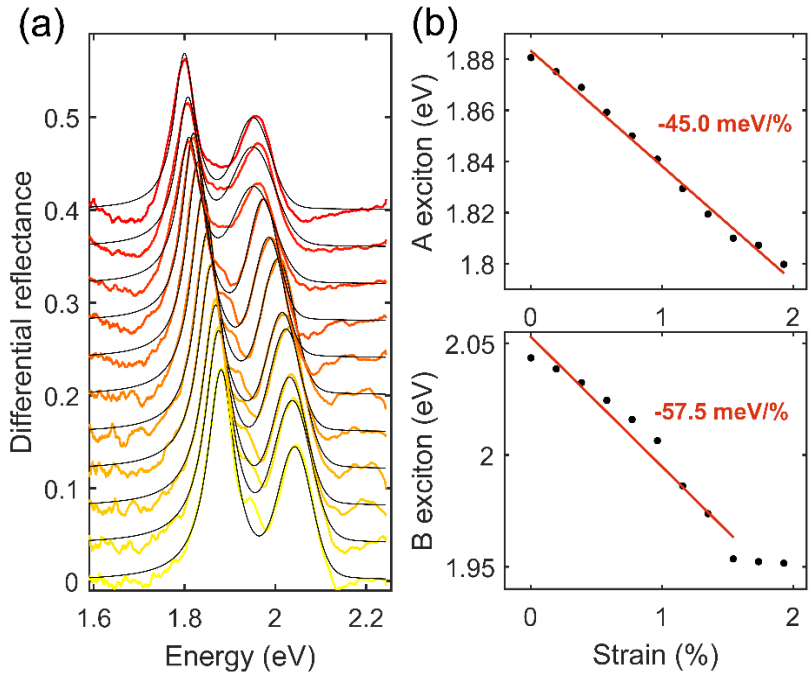

**Figure S2.** Strain tunable differential reflectance in 2L-MoS$_2$ on a polypropylene substrate (185 μm thick). (a) Differential reflectance spectra acquired at different uniaxial strain levels, slippage doesn't occur up to ~2%. (b) A and B exciton energy values as a function of the applied uniaxial strain. A linear fit is employed to extract the gauge factor, the spectral shift per % of uniaxial strain, which is indicated in each panel.

**REPRODUCIBILITY OF THE STRAIN CYCLES IN 3L-MOS$_2$:**

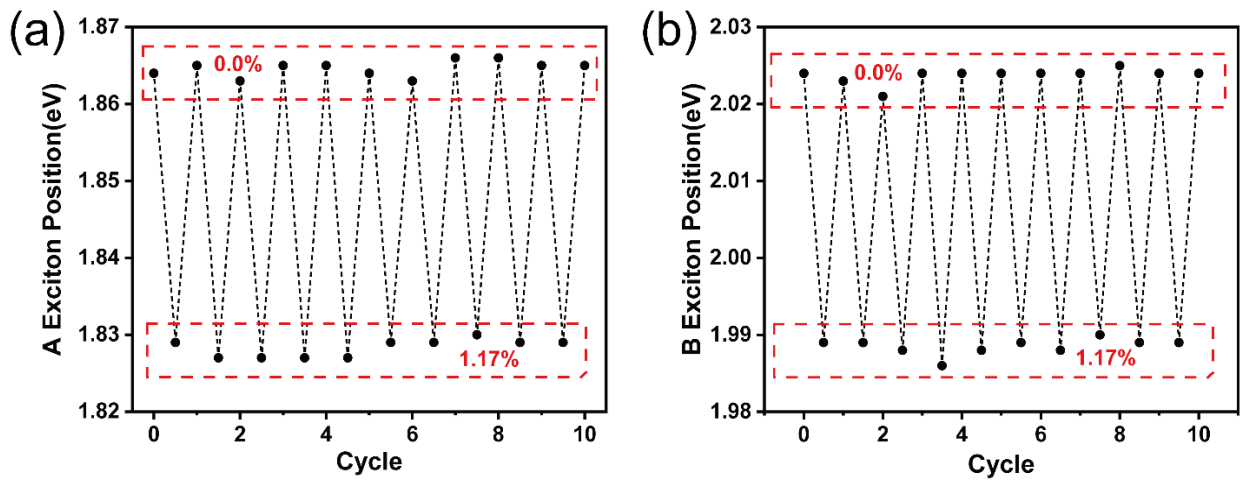

**Figure S3.** Reproducibility of the strain cycles in a 3L-MoS$_2$ (on 750 μm thick PC substrate). (a) and (b) A and B exciton energy values for the different strain cycles between 0% and 1.17% uniaxial strain.

**COMPARISON BETWEEN PHOTOLUMINESCENCE AND DIFFERENTIAL REFLECTANCE IN 1L MOS$_2$:**

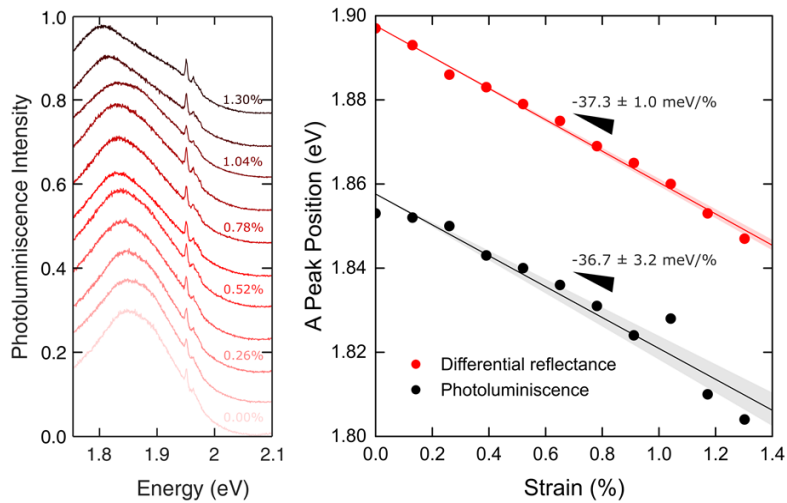

**Figure S4.** Photoluminescence spectra for 1L-MoS$_2$ (on 250 μm thick PC substrate) as a function of uniaxial strain. (left) Photoluminiscence spectra measured at different strain levels. (right) A exciton energy extracted from photoluminescence and from differential reflectance measurements. We attribute the overall offset in the A exciton energy in the PL measurements to the large background signal arising from the substrate that can artificially shift slightly the position of the peaks in the fit process.

## STATISTICAL ANALYSIS OF THE FLAKE-TO-FLAKE VARIATION IN 1L MOS$_2$:

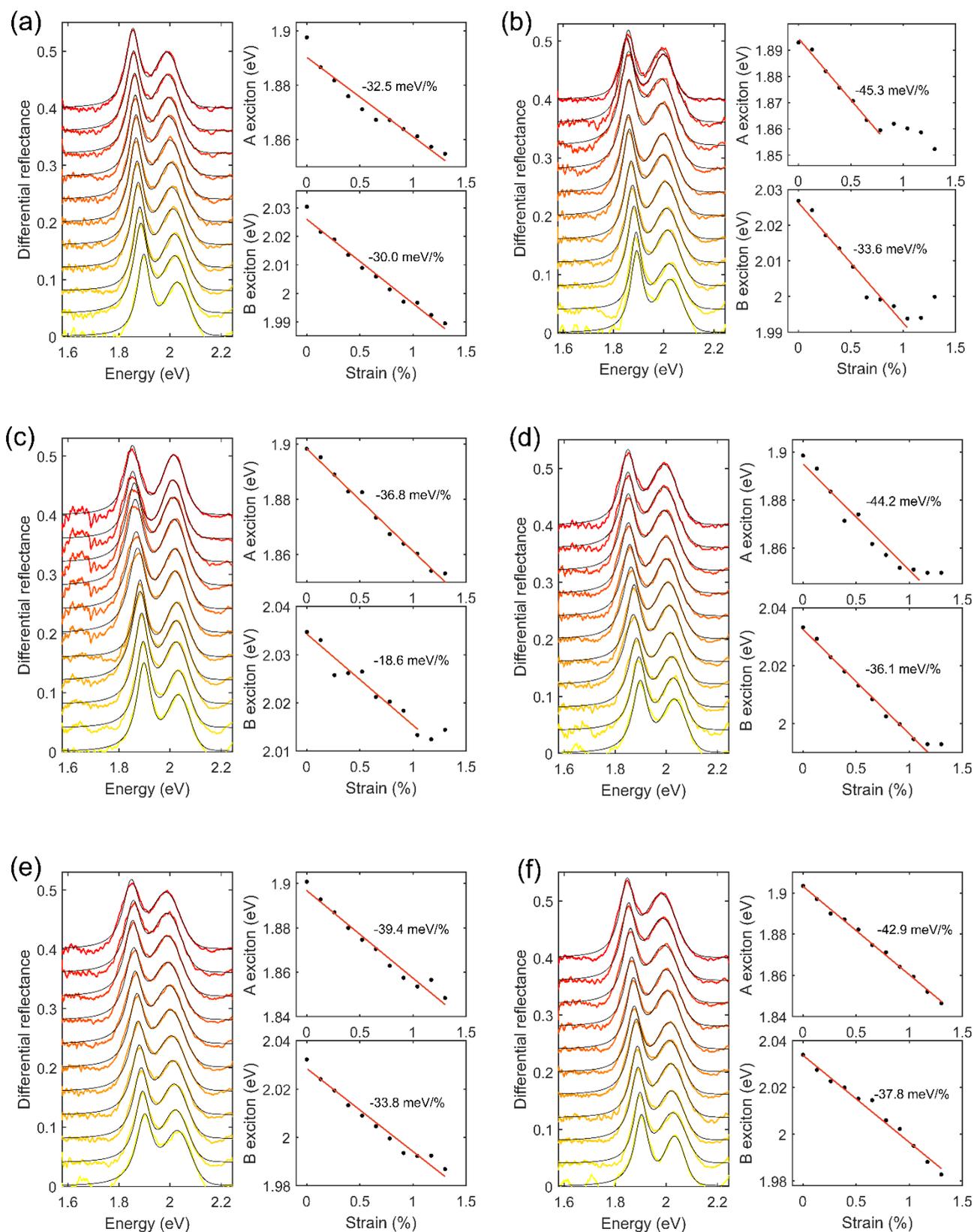

**Figure S5.** Strain tunable differential reflectance in 14 different 1L-MoS$_2$ flakes.

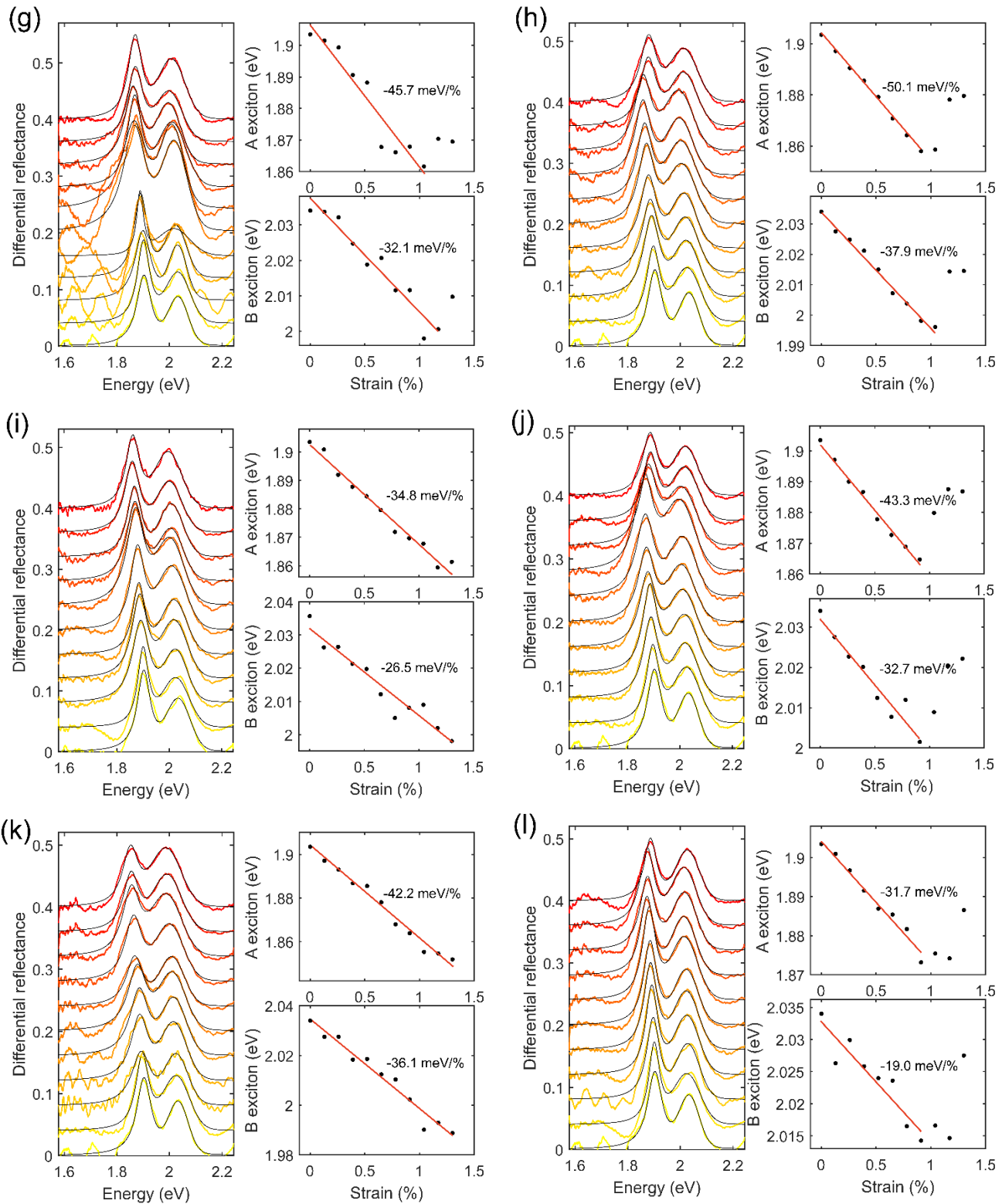

**Figure S5.** Strain tunable differential reflectance in 14 different 1L-MoS$_2$ flakes.

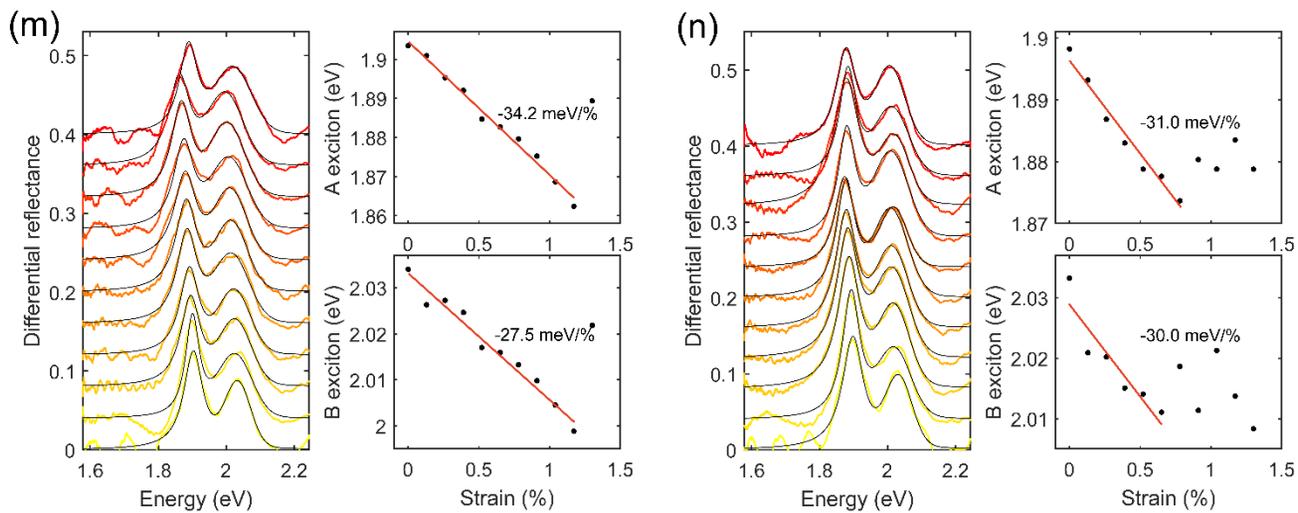

**Figure S5.** Strain tunable differential reflectance in 14 different 1L-MoS$_2$ flakes.

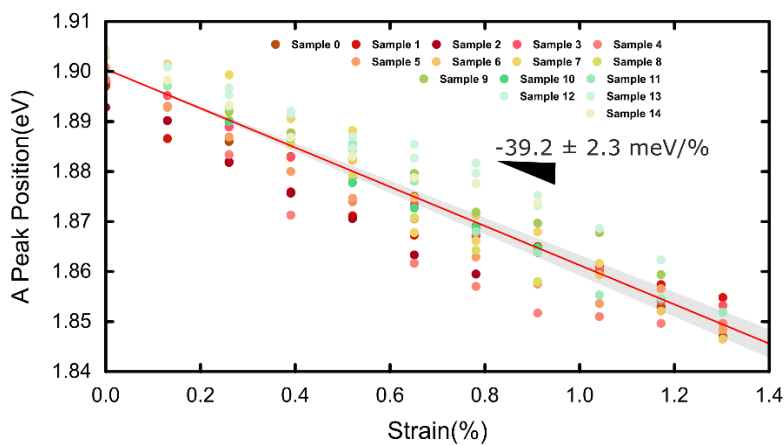

**Figure S6.** Summary of the uniaxial strain engineering measurements on 15 1L-MoS$_2$ flakes. A exciton energy vs. applied uniaxial strain.

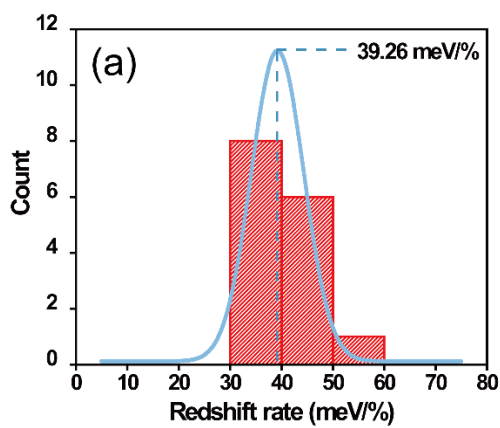

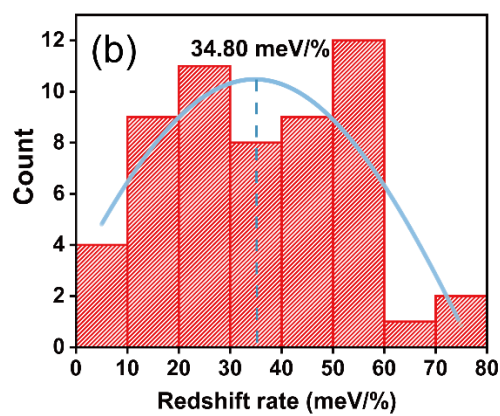

**Figure S7.** Statistical information about the flake-to-flake variation. (a) Histogram of the gauge factor measured on the different 15 1L-MoS$_2$ flakes. (b) Histogram reported in the Supporting Information of reference [8].

**DATASETS FOR 2L AND 3L MOS$_2$:**

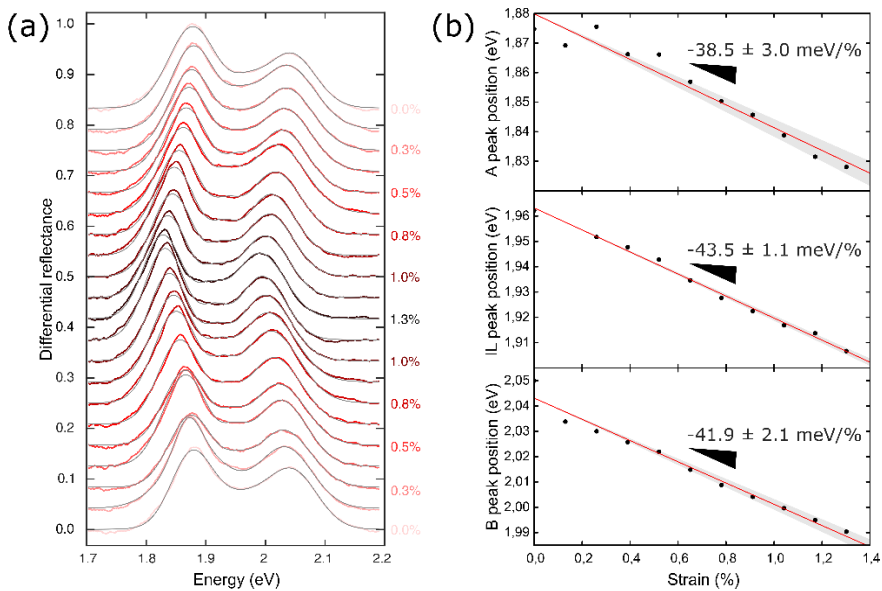

**Figure S8.** Strain tunable differential reflectance in 2L-MoS$_2$. (a) Differential reflectance spectra acquired at different uniaxial strain levels, up to 1.3%. (b) A and B exciton energy values as a function of the applied uniaxial strain. A linear fit is employed to extract the gauge factor, the spectral shift per % of uniaxial strain, which is indicated in each panel.

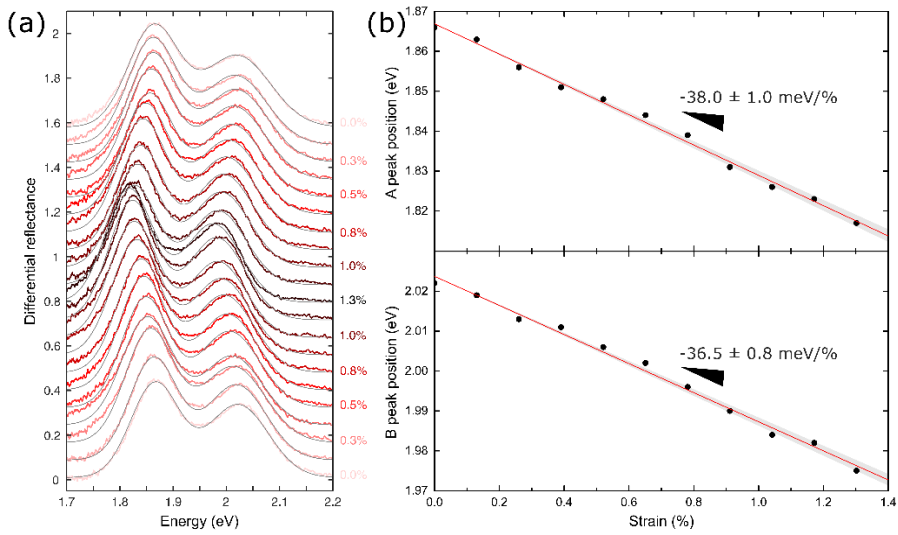

**Figure S9.** Strain tunable differential reflectance in 3L-MoS$_2$. (a) Differential reflectance spectra acquired at different uniaxial strain levels, up to 1.3%. (b) A and B exciton energy values as a function of the applied uniaxial strain. A linear fit is employed to extract the gauge factor, the spectral shift per % of uniaxial strain, which is indicated in each panel.

**DATASETS FOR 1L, 2L AND 3L MOSE$_2$:**

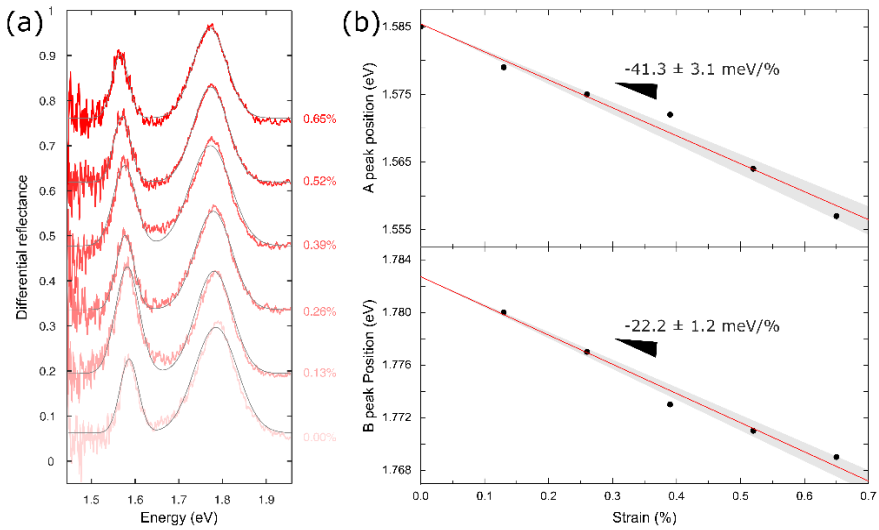

**Figure S10.** Strain tunable differential reflectance in 1L-MoSe$_2$. (a) Differential reflectance spectra acquired at different uniaxial strain levels, up to 0.65%. (b) A and B exciton energy values as a function of the applied uniaxial strain. A linear fit is employed to extract the gauge factor, the spectral shift per % of uniaxial strain, which is indicated in each panel.

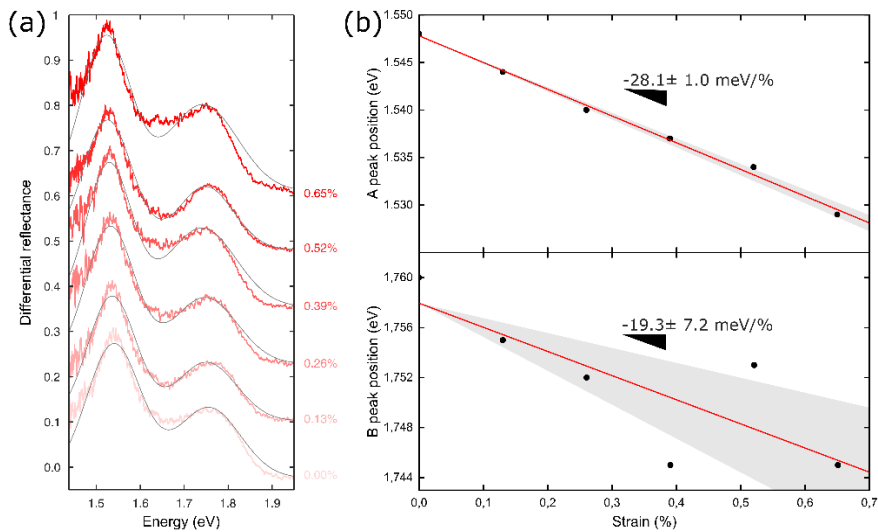

**Figure S11.** Strain tunable differential reflectance in 2L-MoSe$_2$. (a) Differential reflectance spectra acquired at different uniaxial strain levels, up to 0.65%. (b) A and B exciton energy values as a function of the applied uniaxial strain. A linear fit is employed to extract the gauge factor, the spectral shift per % of uniaxial strain, which is indicated in each panel.

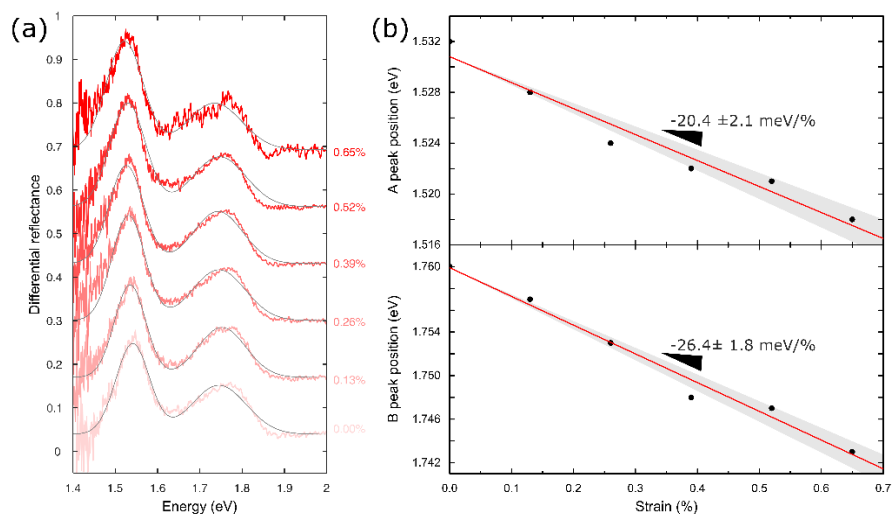

**Figure S12.** Strain tunable differential reflectance in 3L-MoSe$_2$. (a) Differential reflectance spectra acquired at different uniaxial strain levels, up to 0.65%. (b) A and B exciton energy values as a function of the applied uniaxial strain. A linear fit is employed to extract the gauge factor, the spectral shift per % of uniaxial strain, which is indicated in each panel.

**DATASETS FOR 1L, 2L AND 3L WS$_2$:**

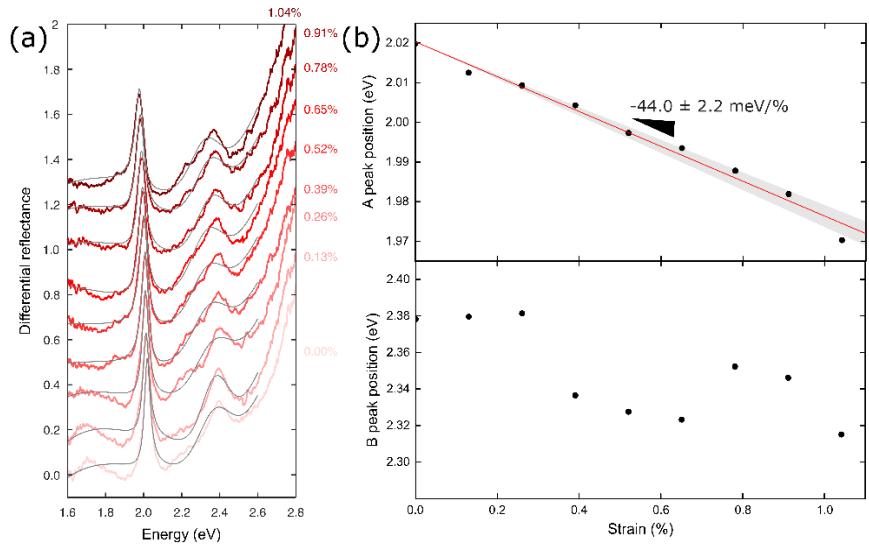

**Figure S13.** Strain tunable differential reflectance in 1L-WS$_2$. (a) Differential reflectance spectra acquired at different uniaxial strain levels, up to 1%. (b) A and B exciton energy values as a function of the applied uniaxial strain. A linear fit is employed to extract the gauge factor, the spectral shift per % of uniaxial strain, which is indicated in each panel.

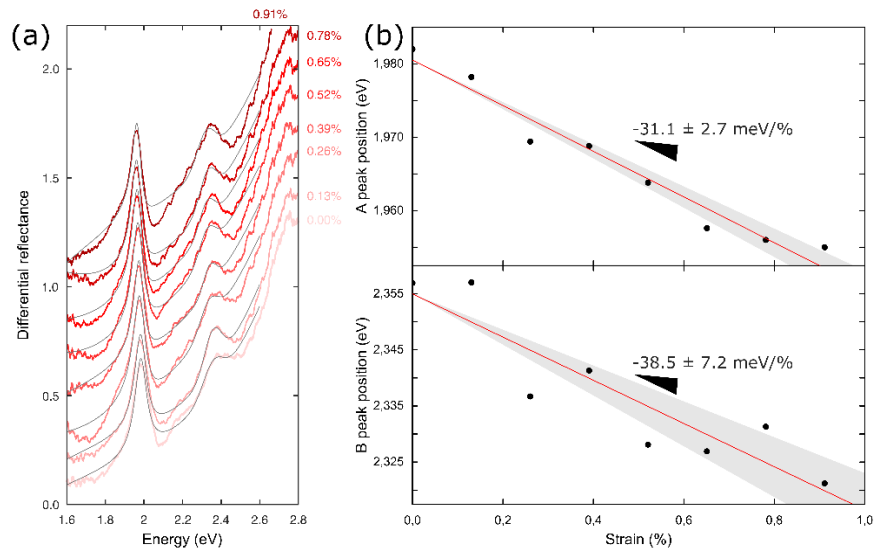

**Figure S14.** Strain tunable differential reflectance in 2L-WS$_2$. (a) Differential reflectance spectra acquired at different uniaxial strain levels, up to 0.9%. (b) A and B exciton energy values as a function of the applied uniaxial strain. A linear fit is employed to extract the gauge factor, the spectral shift per % of uniaxial strain, which is indicated in each panel.

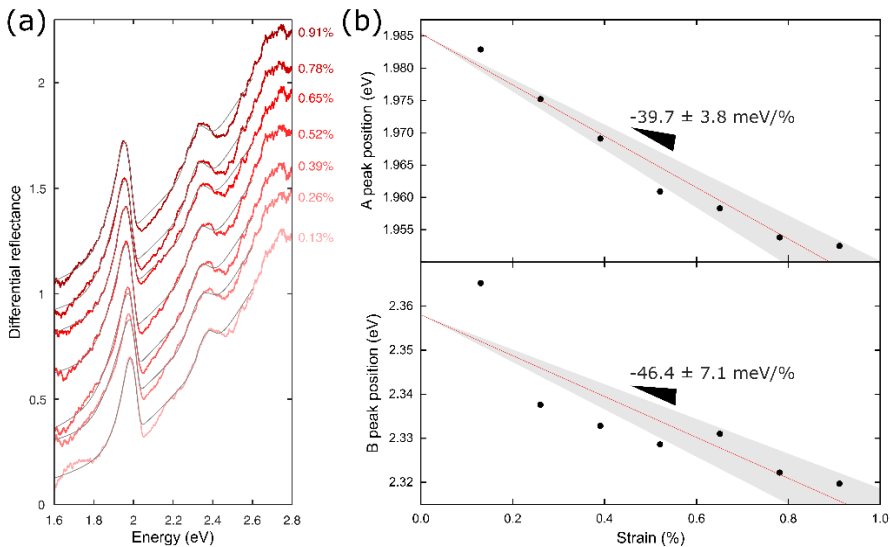

**Figure S15.** Strain tunable differential reflectance in 3L-WS$_2$. (a) Differential reflectance spectra acquired at different uniaxial strain levels, up to 0.9%. (b) A and B exciton energy values as a function of the applied uniaxial strain. A linear fit is employed to extract the gauge factor, the spectral shift per % of uniaxial strain, which is indicated in each panel.

**DATASETS FOR 1L, 2L AND 3L WSE$_2$:**

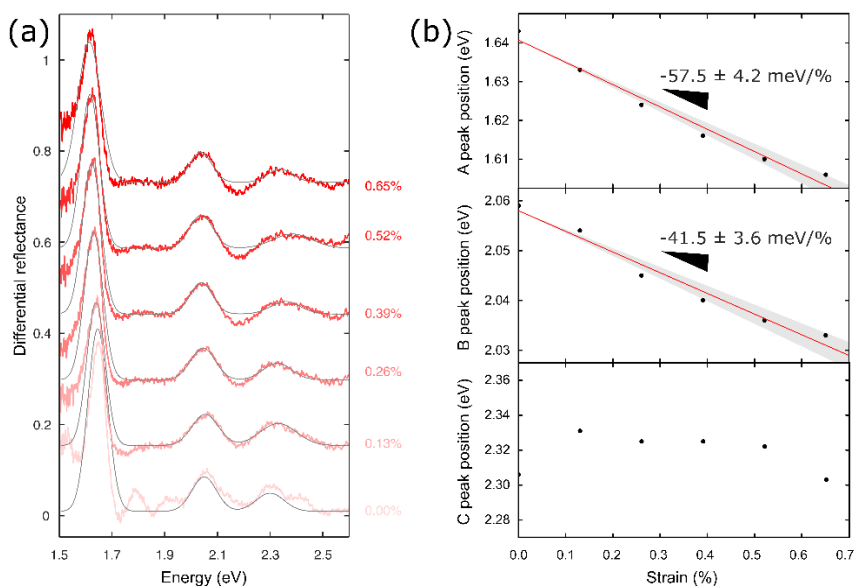

**Figure S16.** Strain tunable differential reflectance in 1L-WSe$_2$. (a) Differential reflectance spectra acquired at different uniaxial strain levels, up to 0.65%. (b) A and B exciton energy values as a function of the applied uniaxial strain. A linear fit is employed to extract the gauge factor, the spectral shift per % of uniaxial strain, which is indicated in each panel.

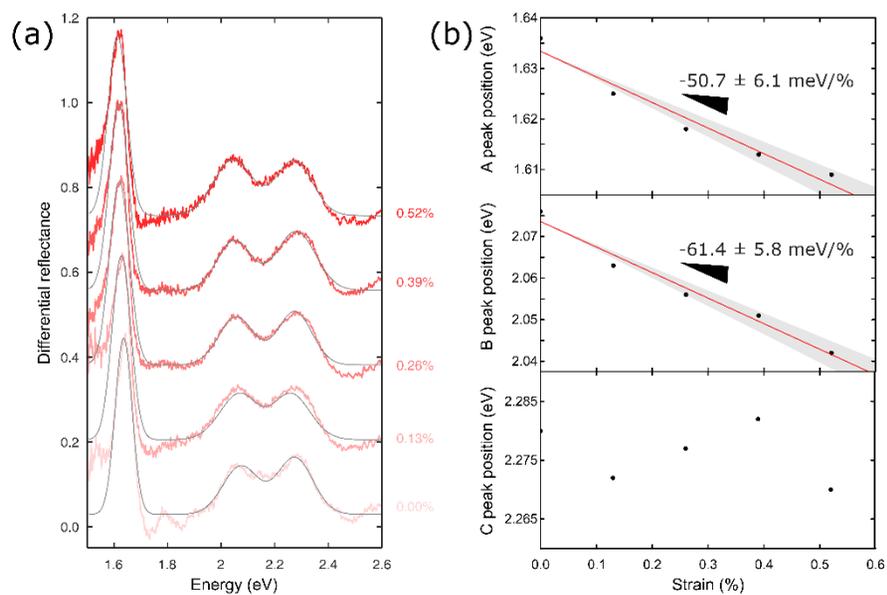

**Figure S17.** Strain tunable differential reflectance in 2L-WSe$_2$. (a) Differential reflectance spectra acquired at different uniaxial strain levels, up to 0.5%. (b) A and B exciton energy values as a function of the applied uniaxial strain. A linear fit is employed to extract the gauge factor, the spectral shift per % of uniaxial strain, which is indicated in each panel.

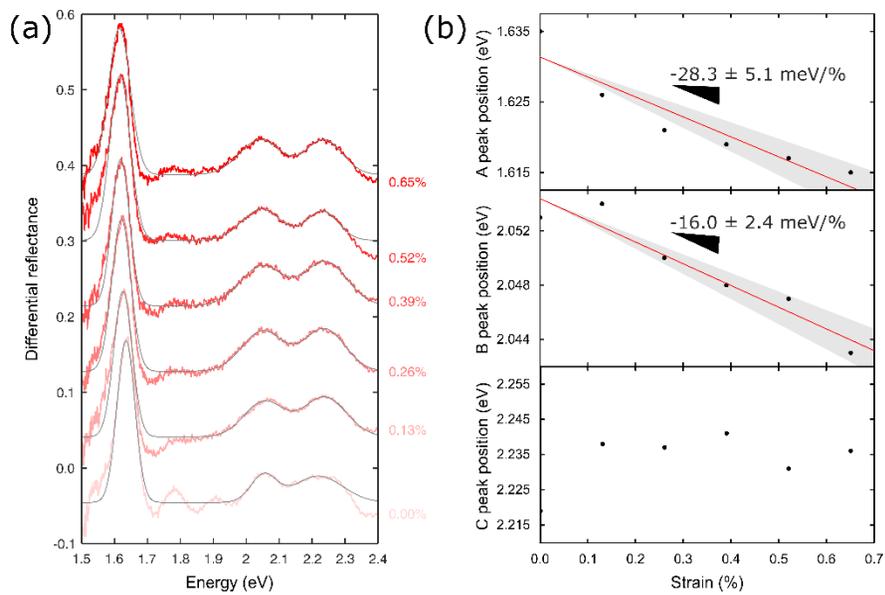

**Figure S18.** Strain tunable differential reflectance in 3L-WSe$_2$. (a) Differential reflectance spectra acquired at different uniaxial strain levels, up to 0.65%. (b) A and B exciton energy values as a function of the applied uniaxial strain. A linear fit is employed to extract the gauge factor, the spectral shift per % of uniaxial strain, which is indicated in each panel.